\documentclass[runningheads]{llncs}
 
\usepackage{eccv}

\usepackage{eccvabbrv}

\usepackage{graphicx}
\usepackage{booktabs}
\usepackage{wrapfig}

\usepackage[accsupp]{axessibility}  
\usepackage{amsmath}

\usepackage{hyperref}

\usepackage{orcidlink}

\DeclareMathOperator*{\argmin}{arg\,min}

\begin{document}
\def\mathbi#1{\textbf{\em #1}}
\title{Refined Geometry-guided Head Avatar Reconstruction from Monocular RGB Video} 

\titlerunning{GeoNeRF}

\author{
Pilseo Park\inst{1}
\and 
Ze Zhang \inst{1,2}
\and
Michel Sarkis \inst{2}\thanks{Formerly of Qualcomm Technologies.}
\and
Ning Bi \inst{2}
\and
Xiaoming Liu \inst{1}
\and
Yiying Tong  \inst{1}\thanks{All of the datasets mentioned in this paper were solely downloaded and used by Michigan State University.}
}

\authorrunning{P. Park et al.}

\institute{Michigan State University
\and
Qualcomm Technologies, Inc.
}

\maketitle

\begin{abstract}
High-fidelity reconstruction of head avatars from monocular videos is highly desirable for virtual human applications, but it remains a challenge in the fields of computer graphics and computer vision. In this paper, we propose a two-phase head avatar reconstruction network that incorporates a refined 3D mesh representation. Our approach, in contrast to existing methods that rely on coarse template-based 3D representations derived from 3DMM, aims to learn a refined mesh representation suitable for a NeRF that captures complex facial nuances. In the first phase, we train 3DMM-stored NeRF with an initial mesh to utilize geometric priors and integrate observations across frames using a consistent set of latent codes. In the second phase, we leverage a novel mesh refinement procedure based on an SDF constructed from the density field of the initial NeRF. To mitigate the typical noise in the NeRF density field without compromising the features of the 3DMM, we employ Laplace smoothing on the displacement field. Subsequently, we apply a second-phase training with these refined meshes, directing the learning process of the network towards capturing intricate facial details. Our experiments demonstrate that our method further enhances the NeRF rendering based on the initial mesh and achieves performance superior to state-of-the-art methods in reconstructing high-fidelity head avatars with such input.

\end{abstract}

\section{Introduction}
\label{sec:introduction}
Reconstructing photorealistic 3D head avatars from monocular videos is crucial in the fields of computer graphics and computer vision, especially for applications such as VR/AR, gaming, and video conferencing. To generate high-fidelity avatar, some 2D-based methods\cite{Siarohin2019AnimatingAO,Siarohin2020FirstOM,Wiles2018X2FaceAN} employ either dense warping fields or sparse motion-specific key points to transfer motion information. By training on large-scale facial video datasets\cite{chung2018voxceleb2,zhu2022celebv} featuring a wide range of identities, these approaches demonstrate the ability to generalize to new faces and can generate satisfactory rendering results from just a single image of any given identity. However, these methods do not impose constraints on the 3D structure of facial geometry, leading to challenges in producing images consistent across multiple viewpoints. Moreover, they tend to produce artifacts when dealing with significantly different expressions or poses.

\begin{figure}[tb]
  \centering
  \includegraphics[width=\textwidth]{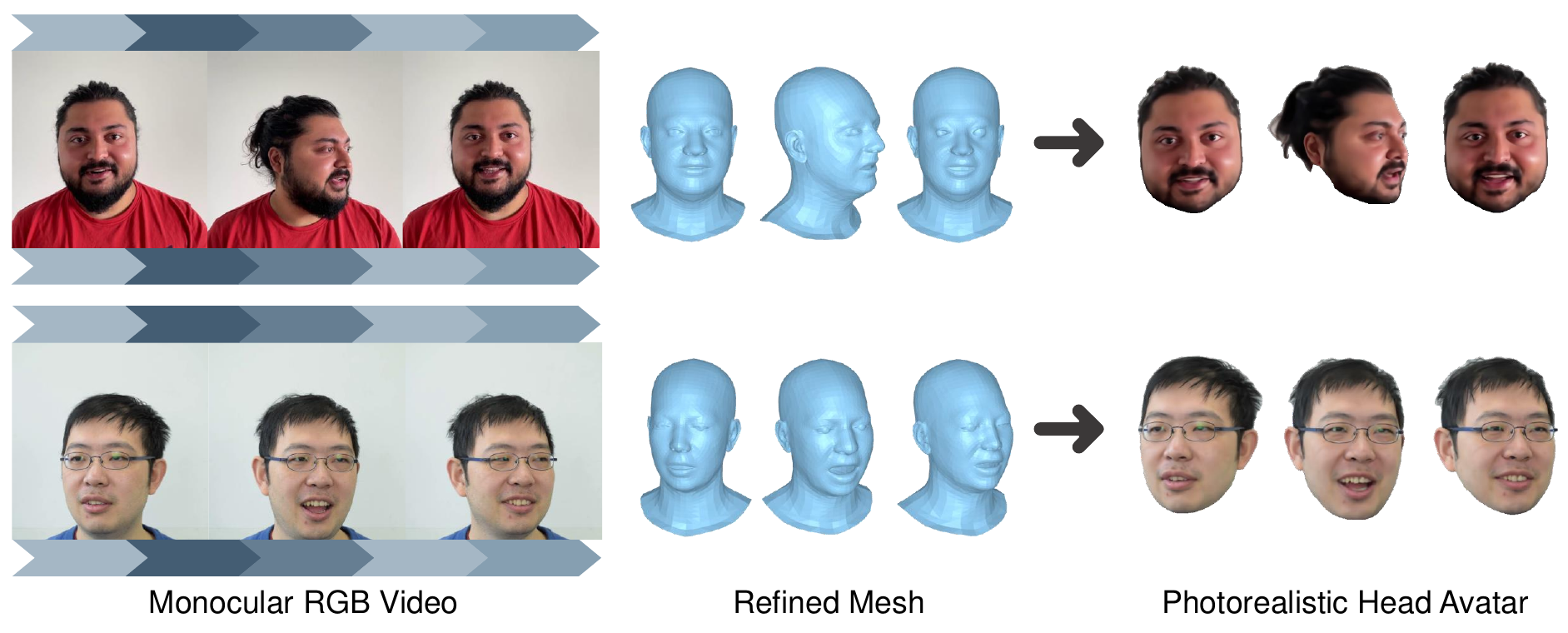}
  \caption{
  Given a monocular RGB video, we train a 3DMM-stored NeRF to generate a refined mesh, which is subsequently used to render a photorealistic head avatar. In contrast to relying on coarse 3DMM template-based geometric representations, our approach captures individualized details, facilitating improved NeRF learning towards refined geometry.
  }
  \label{fig1}
\end{figure}

Recently, Neural Radiance Fields (NeRF)\cite{Mildenhall2020NeRFRS} have demonstrated remarkable capabilities in synthesizing photorealistic renderings from novel viewpoints within 3D scenes. The core idea of NeRF involves utilizing a neural network to represent color and density fields in a volume retrievable from a position in 3D space with a viewing direction, employing volume rendering techniques for novel view synthesis. These realistic rendering abilities have fueled a burgeoning interest in NeRF-based head avatar reconstruction\cite{Gafni2021DynamicNR,Gao2022ReconstructingPS,Athar2022RigNeRFFC,Chen2023ImplicitNH,Bai2023LearningPH}. They achieve facial reconstruction by learning implicit deformation fields or rendering functions conditioned on the expression and pose coefficients from 3D Morphable Models (3DMM)~\cite{Blanz1999AMM}, thus providing a guidance in NeRF-based approaches. 
Despite these significant progresses, the geometric details captured by these methods remain relatively generic due to the sparse viewpoints provided by monocular videos and reliance on predicting the density field using template-based meshes from 3DMM.

In this work, we propose a 3DMM-informed NeRF guided by mesh refinement to address the aforementioned limitations of the existing methods. 
Our work is fundamentally motivated by a pivotal question: \emph{How should we define the geometry that can effectively facilitate the learning of NeRF?} 
In other words, what kind of geometric structure would best support the neural network training? 
We aim to explore a possible answer to this question within NeRF itself, by utilizing the potential of its density field. 
There are three main challenges to be addressed in order to achieve this: 
\textbf{1)} determining how the existing template mesh (geometry) should be further specialized for an individual, 
\textbf{2)} overcoming the restriction in viewing angle from each frame to utilize the entire density field, and 
\textbf{3)} managing the inherent noisiness of the density field of a NeRF trained from sparse input. 

To tackle these issues, we first train the 3DMM-stored NeRF with a template mesh, termed as phase one training for bootstrapping the refinement. 
Leveraging the density field offered by NeRF of a 3D scene, we create a 2.5D mesh (height map) delineating the facial surface of the NeRF. 
We then construct a Signed Distance Function (SDF) to facilitate the signed distance query from a spatial point.
Next, we cast a ray toward each vertex from an outside point along the normal direction, to find the SDF value representing the surface at each sample point along the ray, and perturb the initial vertex location to the surface point as indicated by the density field. 

The resulting mesh perturbation can still be noisy with a varying degree depending on the angle, since the density field used is primarily from a single view from each frame. As our confidence in the density fields is higher for frontal views, we strategically adjust the perturbation to focus mainly on the face's frontal parts by restricting the perturbation range based on the alignment of the surface normal and the viewing direction. Moreover, we employ implicit Laplacian smoothing~\cite{Meyer2002DiscreteDO,SorkineHornung2005LaplacianMP} to remove high-frequency noise without overly flattening the features. 
To prevent volume reduction that may lead to the loss of facial geometric details during such smoothing, we apply Laplacian smoothing to the displacement between the initial and perturbed meshes. 
With this refined mesh, we proceed to the second phase of NeRF training, which enhances the learning of NeRF towards refined geometry.

In summary, our main contributions are as follows:
\begin{itemize}
    \item A novel mesh perturbation algorithm that utilizes SDF constructed from NeRF's density field, focusing primarily on the head's frontal view to guide the reconstruction of a more accurate NeRF.
    \item Introduction of displacement-only implicit Laplacian smoothing that prevents smoothing out the high curvature regions in the phase-one mesh, thereby preserving the facial geometric details.
    \item Experimental studies on head avatar reconstruction that demonstrate our method outperforms state-of-the-art methods in capturing detailed expressions and facial structures.
\end{itemize}

\section{Related Work}
\label{sec:rw}
\subsection{Monocular 3D Head Avatar Reconstruction}
The field of 3D facial modeling and head avatar reconstruction is a dynamic and growing area of research interest. Recognizing the potential of NeRF\cite{Mildenhall2020NeRFRS} for producing photorealistic renderings, a variety of methods\cite{Gafni2021DynamicNR,Athar2022RigNeRFFC,Yu2023NOFANO,Zhao2023HAvatarHH} have been proposed to create controllable 3D head avatars. These methods typically involve conditioning NeRF with global expression parameters extracted from 3D Morphable Models (3DMM) tracking\cite{Blanz1999AMM}. For instance, NerFace\cite{Gafni2021DynamicNR} directly integrates 3DMM-derived expression parameters into a dynamic NeRF, enabling the reconstruction of expressive and controllable 3D facial avatars. RigNeRF\cite{Athar2022RigNeRFFC} employs per-point deformation in a radiance field to adjust head pose and facial expressions, deforming rays into a canonical space defined by 3DMM for color sampling. Moreover, the combination of voxel fields at multiple levels with expression coefficients in the latent space has made it possible to accurately represent head geometry through multi-resolution features~\cite{Gao2022ReconstructingPS}. A line of works\cite{Khakhulin2022RealisticOM,Zheng2022IMA,Zheng2023PointAvatarDP, Zielonka2023InstantVH,Xu2023AvatarMAVF3,Duan2023BakedAvatarBN} have explored integrating traditional 3DMM approaches with neural rendering, using the 3DMM face template as a guide for deformation. IMAvatar\cite{Zheng2022IMA} utilizes neural implicit surfaces\cite{Mescheder2019OccupancyNL,Park2019DeepSDFLC}  to construct an implicit deformation field that transitions from canonical to observed space, guided by expression and pose parameters. PointAvatar\cite{Zheng2023PointAvatarDP} introduces a deformable point-based model, rendering each point as a 2D circle with a uniform radius, and enables the transformation of the point cloud into new poses using expression and pose parameters. Despite substantial progress, the geometry captured by these methods remains relatively coarse because they depend on estimating head geometry using template-based meshes from 3DMM. Instead of solely relying on coarse template-based geometric representations, our approach introduces a refined geometric representation through mesh deformation for the photorealistic rendering of 3D head avatars through a second-phase NeRF.

\subsection{Implicit Geometry-stored Representation}
Storing sparse feature representations within geometry has been shown to improve the visual fidelity of neural radiance fields\cite{liu2020neural,Peng2021NeuralBI,yang2022neumesh,Bai2023LearningPH,Zhuang2023DreamEditorT3}. Neural Body~\cite{Peng2021NeuralBI} embeds latent codes within a vertex-based human body model (SMPL)\cite{loper2015smpl} to control their spatial locations. Furthermore, NSVF\cite{liu2020neural} attaches a voxel embedding to each vertex within a sparse voxel octree structure, and the representation of a query point inside the voxel is obtained by aggregating the voxel embeddings at the vertices of the corresponding voxel. MonoAvatar\cite{Bai2023LearningPH} uses a 3DMM-anchored neural radiance field, and decodes features attached to vertices by finding the k-nearest neighbors for a given 3D query point. These geometry-stored representations are interpolated to decode the neural radiance field during volume rendering, implying that any modification to the mesh geometry will be reflected in the rendering result\cite{yang2022neumesh}. We utilize this representation to form a head geometric prior by perturbing the mesh vertices for surface location determination based on the NeRF density field itself.

\subsection{Mesh Deformation}
\begin{sloppypar}
Meshes, with their efficiency and versatility, facilitate the direct representation of complex shapes and structures through deformation\cite{botsch2010polygon,Shirman1987LocalSI}. Generally, mesh deformation can be divided into geometry-based and data-driven methods. Geometry-based methods\cite{SorkineHornung2005LaplacianMP,Yu2004MeshEW,Au2006DualLE,SorkineHornung2007AsrigidaspossibleSM} address mesh deformation by formulating it as a constrained optimization problem. However, these geometry-based methods fail to accurately model how objects deform and can lead to unintended results, especially for intricate shapes or extensive deformations\cite{Gao2017SparseDD}. Since human faces show substantial deformation with various poses and expressions, we avoid extensive deformation by starting from a template-based 3DMM that incorporates the pose and expression variation, and then adopting Laplacian smoothing\cite{SorkineHornung2005LaplacianMP} to mitigate the noise in the perturbation based on the density field of NeRF. Data-driven methods\cite{Gao2016EfficientAF,Gao2017SparseDD} investigate common deformations in the dataset and thus generate more realistic results. Recently, the proliferation of extensive datasets has facilitated the integration of implicit neural networks into the domain of 3D mesh modeling\cite{Deng2021DeformedIF,Zhang2023SelfsupervisedLO,Shuai2023DPFNetCE,liao2024multi}. These methods investigate 3D modeling on an implicit neural representation for generic objects such as cars, chairs, and the human body and extract the mesh using Marching Cubes~\cite{Lorensen1987MarchingCA}. In contrast to these works, we consider the geometric deformation of faces targeted for a better NeRF and utilize ray marching for each initial vertex to perturb the mesh based on a normal offset leveraging a signed distance function created from a neural radiance field.
\end{sloppypar}

\begin{figure}[tb]
  \centering
  \includegraphics[width=\textwidth]{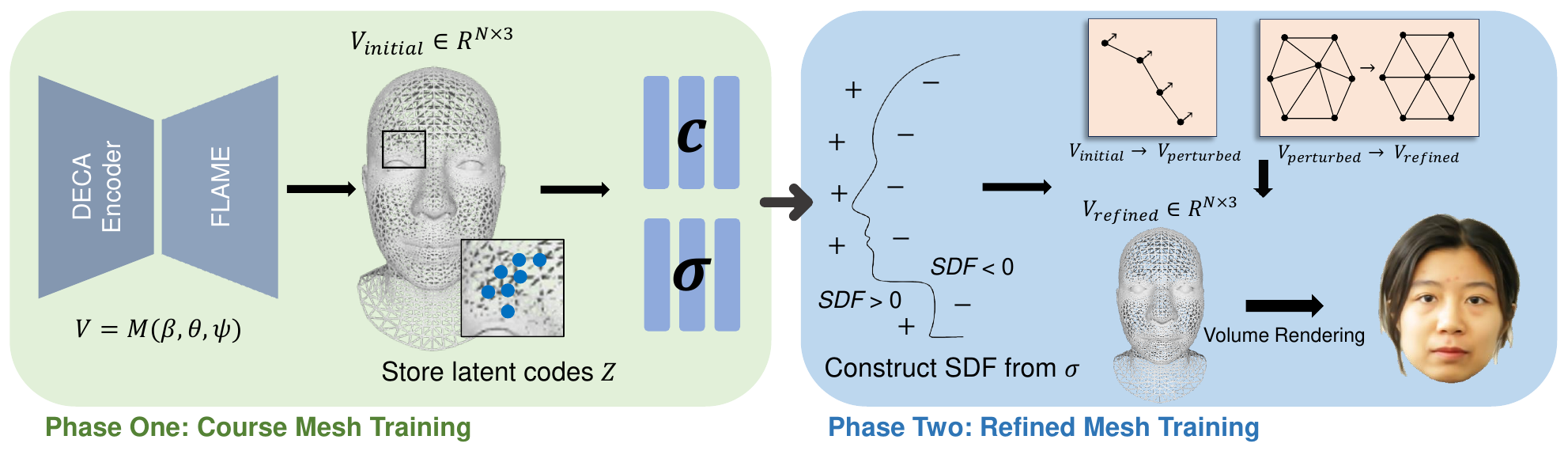}
  \caption{
  Framework overview. In phase one, we obtain one initial template mesh per frame from a pre-trained DECA. We then attach latent codes on mesh vertices to represent the local appearance and geometry of the head, and train the NeRF. In phase two, we construct an SDF from the trained NeRF model and use its density field to refine the mesh through denoised perturbation. Subsequently, we apply second-phase NeRF training and employ volume rendering to generate the final head avatar.
  }
  \label{fig2}
\end{figure}

\section{Methodology}
\label{sec:method}

\subsection{Model Overview}
\label{subsec:overview}
In the context of our work, the goal is to accurately reconstruct and render a 3D model of a human head from a monocular RGB video containing $K$ frames \{${F}_{1}, {F}_{2}, \ldots, {F}_{K}$\}, with the guidance of geometric information. As illustrated in Fig.~\ref{fig2}, our framework consists of two phases: 1) Coarse mesh training, where we train a 3DMM-stored Neural Radiance Field (NeRF) model using the coarse 3D mesh of the face; and 2) Refined mesh training, where we construct Signed Distance Function (SDF) values from the trained NeRF model, refine the mesh and apply second-phase training. 
In the first phase, we utilize DECA~\cite{Feng2021LearningAA} to obtain the initial vertex coordinates ${V}_{\text{initial}}$ $\in$ $\mathbb{R}^{N\times3}$ of the FLAME model~\cite{Li2017LearningAM} (Sec. \ref{subsec:mesh}) for each frame, where $N$ is the number of vertices. Then, we attach a set of latent codes to the vertices to represent a human head (Sec.~\ref{subsec:latent}), and train the NeRF model similar to \cite{Peng2021NeuralBI}. In the second phase, we refine the vertices of the initial mesh by utilizing the SDF values to perturb the vertices towards the surface geometry of the head within the density field, following the typical ray marching (Sec. \ref{subsec:perturbation}). Additionally, we apply Laplacian smoothing\cite{Meyer2002DiscreteDO,SorkineHornung2005LaplacianMP} to mitigate the noise resulting from the mesh perturbation (Sec. \ref{subsec:laplace}). Then we train the model with the refined mesh. This enables the refined meshes to guide the NeRF’s learning process, focusing on capturing fine details. We subsequently employ volumetric rendering to produce novel views of the head avatar (Sec. \ref{subsec:rendering}) using pose and expression parameters as input. More specifically, the input pose and expression parameters induce a 3DMM, which is refined by the phase one NeRF to define the phase two NeRF. The proposed method is trained by minimizing the difference between the rendered images and input images (Sec.~\ref{subsec:objective}).

\subsection{Initial Head Mesh Reconstruction}
\label{subsec:mesh}
Before training our first-stage NeRF, we estimate the FLAME parameters and camera parameter $c$ for each frame using DECA \cite{Feng2021LearningAA}.  FLAME\cite{Li2017LearningAM} is a vertex-based parametric 3D head model that employs  Linear Blend Skinning (LBS) and consists of $N =$ 5023 vertices. It includes a mean template ${V}_{\mathrm{template}} \in \mathbb{R}^{N\times3}$ with blend skinning function $W$, shape blendshapes ${B}_{S}(\beta;\mathcal{S}) : \mathbb{R}^{|\beta|}\rightarrow\mathbb{R}^{N\times3}$, pose correctives ${B}_{P}(\theta;\mathcal{P}) : \mathbb{R}^{|\theta|}\rightarrow\mathbb{R}^{N\times3}$, and expression blendshapes ${B}_{E}(\psi;\mathcal{E}) : \mathbb{R}^{|\psi|}\rightarrow\mathbb{R}^{N\times3}$. These blendshapes are constructed from an extensive dataset of 4D human head scans, enabling FLAME to effectively model diverse facial variations. The mesh in world coordinates is defined as: 
\begin{equation}
  M(\beta,\theta,\psi) = W({V}_{\mathrm{template}} + {B}_{S}(\beta;\mathcal{S}) + {B}_{P}(\theta;\mathcal{P}) + {B}_{E}(\psi;\mathcal{E})),
  \label{eq:eq1}
\end{equation}
where the parameter $\beta \in \mathbb{R}^{|\beta|}$ denotes facial identity, $\theta \in \mathbb{R}^{3f+3}$ (with $f$ = 4 joints for neck, jaw, and eyeballs) denotes pose, and $\psi \in \mathbb{R}^{|\psi|}$ denotes expression, respectively. We denote the model in world coordinates for the $k$-th frame by 
\begin{equation}
{V}_{k} = M({\beta}_{k},{\theta}_{k},{\psi}_{k}) \in \mathbb{R}^{N\times3}.
\label{eq:eq1123}
\end{equation}

\subsection{Geometry Representation}
\label{subsec:latent}
Following the structure of geometry-stored representation-based NeRF\cite{Peng2021NeuralBI}, we attach latent codes $\mathcal{Z} = \{{z}^{1}, {z}^{2}, \ldots, {z}^{N}\}$ to each vertex of the initial mesh ${V}_{k}^{n} = \{{v}_{k}^{1}, {v}_{k}^{2} \ldots, {v}_{k}^{N}\}$. This attachment to a deformable model facilitates the representation of the local geometry of the head and enhances the controllability of the spatial locations of the latent codes. Additionally, by assigning the same set of latent codes across different frames, we effectively integrate observations to tackle the challenge of reconstructing the head from sparse views.

We utilize the embedding of the deformable model to query the latent codes within a bounding volume of the head, allowing multilayer perceptrons (MLPs) to accurately predict the density and color for each point in the implicit field.  However, the structured latent codes are sparsely distributed in volumetric space, and direct interpolation leads to zero vectors at most points. To address this issue, we diffuse the sparse latent codes into the bounding volume. Similar to \cite{Shi2020PVRCNNPF,Peng2021NeuralBI}, we use Submanifold Sparse Convolutional Networks (SSCNs)\cite{Graham20183DSS} to efficiently process the structured latent codes. Through convolution and downsampling, we transform sparse codes into dense codes, resulting in a dense latent code volume. Consequently, we can query the latent code of any point $\mathbf{q}$ from this dense volume. The latent code at query point $\mathbf{q}$ is represented as $\mathcal{G}(\mathbf{q}, \mathcal{Z}, {V}_{k})$. Subsequently, we train the local feature-based neural radiance field, a process we describe as phase one training.

\begin{figure}[tb]
  \centering
  \begin{subfigure}{.6\textwidth}
    \centering
    \includegraphics[width=\linewidth]{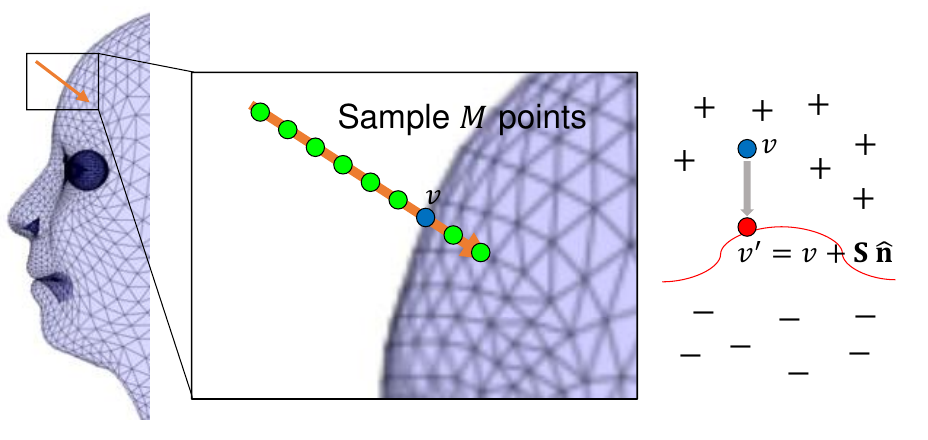}
    \caption{Mesh Perturbation}
    \label{fig3-a}
  \end{subfigure}%
  \begin{subfigure}{.4\textwidth}
    \centering
    \includegraphics[width=\linewidth]{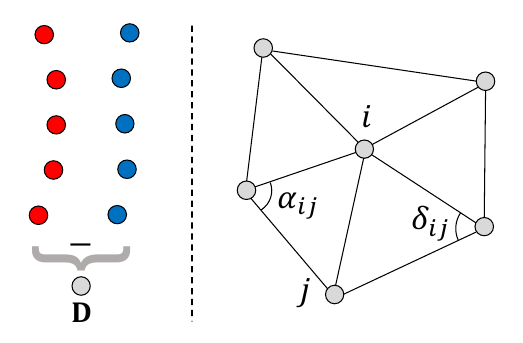}
    \caption{Laplacian Smoothing}
    \label{fig3-b}
  \end{subfigure}
  \caption{ Denoised mesh perturbation.
  \textbf{(a)} From $M$ sample points along the ray, we identify the displacement amount $\mathbf{S}$, which represents the distance to the surface from the initial vertex $v$. We then perturb $v$ in the direction of the normal $\mathbf{\hat{n}}$ using $\mathbf{S}$ to obtain $v'$. \textbf{(b)} To apply displacement-only smoothing, we first obtain the displacement vector field $\mathbf{D}$. For vertex $i$ and its neighbor $j$, $\alpha_{ij}$ and $\delta_{ij}$ represent the angles opposite the edge between them, which are used to assemble the geometric Laplacian matrix $\mathbf{L}$.
  }
  \label{fig3}
\end{figure}

\subsection{Mesh Perturbation}
\label{subsec:perturbation}
The initial mesh obtained in FLAME format~\cite{Li2017LearningAM} provides us with the coarse geometry of the human head. To capture better aligned NeRF, we refine the mesh by perturbing the vertex locations while keeping the same connectivity (see Fig. \ref{fig3-a}). As NeRF models a 3D scene as fields of volume density and colors, we construct a 2.5D mesh from this density field, to represent a less noisy version of the surface than any of the level sets of the density field. We then convert the 2.5D surface (height function) into a signed distance function (SDF) for a simpler implementation of mesh deformation based on offsetting vertices along the local normal direction. We adapted a typical ray marching procedure by constructing rays that pass through outward offset initial mesh vertices, directed inward along the normal at each vertex. 

Specifically, our confidence in density fields is higher for near frontal views due to predominantly frontal sources in typical training sets, which complicates accurate prediction of the back of the head. Therefore, we adjust the ray's initialization point and calibrate its maximum displacement magnitude using $\text{ReLU}\left(\langle \mathbf{\hat{n}}, \hat{z} \rangle\right)$, where $\mathbf{\hat{n}}$ is the unit normal vector of each vertex and $\hat{z}$ is the canonical $z$-axis unit vector. This ensures more adjustments are allowed when the normal is more aligned with the positive $z$-direction. We then sample $M$ points along each ray, computing the SDF value at each to determine the distance to the nearest surface point. The perturbation magnitude $\mathbf{S}$ to the surface is determined by identifying the minimum distance ${t}_{m}$ at which the SDF value reaches minimum magnitude, indicating proximity to the surface:
\begin{equation}
\mathbf{S} = \argmin_{\{t_m|m=1,\dots,M\}} |\text{SDF}({V}_{\mathrm{initial}} + {t}_{m} \, \mathbf{\hat{n}})|.
\label{eq:eq112341}
\end{equation}
If SDF at the perturbed location has a magnitude above a threshold $\epsilon$, we consider the density field noisy for that vertex and set $\mathbf{S}$ for the vertex to 0. Then, the perturbed vertex position ${V}_{\mathrm{perturbed}}$ is computed by adjusting the initial vertex position ${V}_\mathrm{initial}$ along the direction of the normal vector $\mathbf{\hat{n}}$ by the distance $\mathbf{S}$:
\begin{equation}
{V}_{\mathrm{perturbed}} = {V}_{\mathrm{initial}} + \mathbf{S} \, \mathbf{\hat{n}}.
\label{eq:eq1123411}
\end{equation}
By perturbing the vertices towards the surface based on the calculated interaction distance $\mathbf{S}$, the mesh is more consistent with the density field of the NeRF.

\subsection{Laplacian Smoothing}
\label{subsec:laplace}
In Sec. \ref{subsec:perturbation}, we discussed the process of mesh perturbation based on constructing SDF from the trained NeRF model. Although NeRF provides a powerful representation for capturing scene geometry, the resulting mesh can exhibit noise and artifacts due to the discrete nature of sampling, even with our treatment of integrating the density field first along the frontal direction to obtain the SDF. To address this issue, we employ cotangent weight Laplacian smoothing\cite{Meyer2002DiscreteDO,SorkineHornung2005LaplacianMP} to ensure achieving a smooth mesh (see Fig. \ref{fig3-b}). First, we construct the Laplacian matrix $\mathbf{L} \in \mathbb{R}^{N\times N}$ employing cotangent weights to accurately capture the local geometric relationships among vertices:
\begin{equation}
\mathbf{L}_{ij} = 
\begin{cases} 
\hfil w_{ij} = - \frac{1}{2}(\cot (\alpha_{ij}) + \cot (\delta_{ij})) & \hfil j \in \mathcal{N}(i) \\
\hfil \sum\limits_{k \in \mathcal{N}(i)} - w_{ik} & \hfil   i = j \\
\hfil 0 & \hfil \text{otherwise},
\end{cases}
\label{eq:eq5-tmp}
\end{equation}
where $\cot (\alpha_{ij})$ and $\cot (\delta_{ij})$ denote the cotangents of the angles opposite to the edge between vertices $i$ and $j$ in the two incident triangles, and $\mathcal{N}(i)$ is the set of vertices directly adjacent to the vertex $i$.

Since the smoothing process averages the positions of vertices in the mesh based on the positions of their neighboring vertices, it is known to reduce the overall volume of the mesh. This effect is most noticeable in areas of the mesh with high curvature or detail, implying that fine geometric details can be diminished. To prevent this, we apply displacement-only smoothing. Specifically, we target the smoothing of the displacement $\mathbf{D} = {V}_\text{perturbed} - {V}_\text{initial}$ rather than the positions of the vertices themselves:
\begin{equation}
(\mathbf{I} + \lambda \mathbf{L}) \mathbf{X} = \mathbf{D},
\label{eq:eq15-tmp}
\end{equation}
where $\mathbf{X}$ represents the smoothed displacements that we aim to solve for, and $\lambda$ is a regularization parameter adjusting the smoothing intensity. The inclusion of $\mathbf{I}$ in the linear system ensures that the implicit Euler update on $\mathbf{X-D}$ is the Laplacian term $-\lambda \mathbf{L} \mathbf{X}$ associated with the unknown $\mathbf{X}$. Finally, the refined mesh after perturbation and smoothing is defined as:
\begin{equation}
{V}_\text{refined} = {V}_\text{initial} + \mathbf{X}.
\label{eq:eq11234111}
\end{equation}

\subsection{Volumetric Rendering}
\label{subsec:rendering}
For the $k$-th frame, the volume density ${\sigma}_{k}$ at query point $\mathbf{q}$ is predicted by the density MLP network ${\mathcal{F}}_{\sigma}$, which exclusively uses the latent code $\mathcal{G}(\mathbf{q}, \mathcal{Z}, {V}_{k})$ Additionally, the color ${\mathbi{c}}_{k}$ at point $\mathbf{q}$ is predicted by the color MLP network ${\mathcal{F}}_{\mathbi{c}}$, using the latent code $\mathcal{G}(\mathbf{q}, \mathcal{Z}, {V}_{k})$, the viewing direction $\mathbf{d}$, the 3D location $\mathbf{q}$, and the per-frame latent embedding ${\ell}_{k}$ which is used to encode the temporally-varying factors as inputs. Both the viewing direction and the 3D location are processed using the same positional encoders ${\gamma}_{\mathbf{d}}$ and ${\gamma}_{\mathbf{q}}$ as described in NeRF\cite{Mildenhall2020NeRFRS}. Thus, the density and color networks are defined as:
\begin{equation}
  {\sigma}_{k}(\mathbf{q}) = {\mathcal{F}}_{\sigma}(\mathcal{G}(\mathbf{q}, \mathcal{Z}, {V}_{k})),
  \label{eq:eq6-tmp}
\end{equation}
\begin{equation}
{\mathbi{c}}_{k}(\mathbf{x}) = \mathcal{F}_{\mathbf{c}}(\mathcal{G}(\mathbf{q}, \mathcal{Z}, {V}_{k}), {\gamma}_{\mathbf{d}}(\mathbf{d}), {\gamma}_{\mathbf{q}}(\mathbf{q}), \ell_{k}).
  \label{eq:eq33-tmp}
\end{equation}

Finally, we adapt the NeRF volume rendering equations to render the output head image. Let $\mathbi{r}(t) = \mathbi{o} + t\mathbf{d}$ be a point along the camera ray, with ${t}_{n}$ and ${t}_{f}$ specifying the near and far bounds on that ray, respectively. Then the expected color $\hat{\mathbi{C}}$ at frame $k$ is given by:
{
\footnotesize
\begin{equation}
\begin{aligned}
\hat{\mathbi{C}_{k}}(\mathbi{r}) = \int_{{t}_{n}}^{{t}_{f}} T(t){\sigma}_{k}(\mathbi{r}(t)){\mathbi{c}}_{k}(\mathbi{r}(t),\mathbf{d}) \,dt, \text{where } T(t) = \exp\left(-\int_{t_n}^{t} \sigma_k(\mathbi{r}(s)) \, ds\right).
\end{aligned}
\label{eq:eq44-tmp}
\end{equation}
}

\subsection{Training Objective}
\label{subsec:objective}
Our model is trained on monocular RGB videos, and we optimize the model using the mean squared error between the rendered and ground truth images. The corresponding loss function is defined as:
\begin{equation}
\mathcal{L} = \sum_{\mathbi{r}\in\mathcal{R}} \left\| \hat{\mathbi{C}}(\mathbi{r}) - {\mathbi{C}}(\mathbi{r})\right\|_2^2,
\label{eq:eq55-tmp}
\end{equation}
where $\mathcal{R}$ is the set of camera rays that intersect the pixels of the image, and $\hat{\mathbi{C}}(\mathbi{r})$ and ${\mathbi{C}}(\mathbi{r})$ are the rendered pixel color and the ground truth color, respectively.

\section{Experiments}
\label{sec:experiments}

\subsection{Datasets}
\label{subsec:datasets}
We evaluate our method for head avatar reconstruction using publicly available videos of 6 subjects, where subjects 1 and 2 are from IMAvatar~\cite{Zheng2022IMA}, subject 3 is from PointAvatar~\cite{Zheng2023PointAvatarDP}, and subjects 4, 5, and 6 are from NerFace~\cite{Gafni2021DynamicNR}. The video datasets from IMAvatar and PointAvatar consist of about 4,000 frames per subject for training and 1,500 frames each for testing. While the training videos primarily consist of neutral expressions, the testing videos include a wider range of unseen and challenging facial expressions, such as jaw drops and broad smiles. The NerFace dataset consists of sequences for each subject, each roughly 6,000 frames long with the last 1,000 frames of each sequence being used to test the reconstruction.

\subsection{Metrics}
\label{subsec:metrics}
To evaluate the quality of reconstructed head avatars, we follow common metrics as outlined in prior work\cite{Gafni2021DynamicNR}: the $L_1$ distance, Peak Signal-to-Noise Ratio (PSNR), Structure Similarity Index (SSIM)\cite{Wang2004ImageQA}, and Learned Perceptual Image Path Similarity (LPIPS)\cite{Zhang2018TheUE}.

\subsection{Implementation Details}
\label{subsec:implementation_details}
We sample $M = 32$ points and the dimensionality of the latent code $z$ is set to 16. We apply 10 iterations for Lapalcian smoothing and the regularization parameter is set to $\lambda = 0.05$. We train our model on a single NVIDIA Tesla V100 GPU. We use the Adam optimizer \cite{Kingma2015AdamAM} with ${\beta}_1 = 0.9$, ${\beta}_2 = 0.999$, and the learning rate is set to $5\cdot{10}^{-4}$.

\begin{table}[b]
    \caption{
    Quantitative comparison with SOTA methods. We show the results for all 6 subjects. Subject 1 and 2 are from IMAvatar\cite{Zheng2022IMA}, and subject 3 is from PointAvatar\cite{Zheng2023PointAvatarDP} dataset, respectively. Subjects 4, 5, and 6 are from the NerFace\cite{Gafni2021DynamicNR} dataset.
    }
\renewcommand{\arraystretch}{1.1}
    \centering
    \resizebox{\textwidth}{!}{%
    \begin{tabular}{|ccccc|cccc|clll|} \hline 
         &  \multicolumn{4}{c|}{Subject 1}&  \multicolumn{4}{c|}{Subject 2}&  \multicolumn{4}{|c|}{Subject 3}\\ \hline 
         &  $L_1\downarrow$ &  PSNR$\uparrow$&  SSIM$\uparrow$&  LPIPS$\downarrow$&  $L_1\downarrow$ &  PSNR$\uparrow$&  SSIM$\uparrow$&  LPIPS$\downarrow$&  $L_1\downarrow$ & PSNR$\uparrow$& SSIM$\uparrow$&LPIPS$\downarrow$\\ \hline
         IMAvatar\cite{Zheng2022IMA}&  0.027&  25.882&  \textbf{0.950}&  0.078&  0.027&  22.891&  0.914&  0.103&  0.014& 24.999& 0.946&0.098\\ 
         PointAvatar\cite{Zheng2023PointAvatarDP}&  \textbf{0.023}&  26.524&  0.940&  \textbf{0.060}&  0.030&  22.322&  0.905&  \textbf{0.070}&  0.017& 24.320& 0.928&\textbf{0.071}\\ 
         Ours&  0.024&  \textbf{26.581}&  0.948&  0.079&  \textbf{0.026}&  \textbf{23.764}&  \textbf{0.915}&  0.078&  \textbf{0.013}& \textbf{27.064}& \textbf{0.950}&0.076\\ \hline \hline 
         &  \multicolumn{4}{c|}{Subject 4}&  \multicolumn{4}{c|}{Subject 5}&  \multicolumn{4}{|c|}{Subject 6}\\ \hline 
         &  $L_1\downarrow$ &  PSNR$\uparrow$&  SSIM$\uparrow$&  LPIPS$\downarrow$&  $L_1\downarrow$ &  PSNR$\uparrow$&  SSIM$\uparrow$&  LPIPS$\downarrow$&  $L_1\downarrow$ & PSNR$\uparrow$& SSIM$\uparrow$&LPIPS$\downarrow$\\ \hline
         NerFace\cite{Gafni2021DynamicNR}&  0.014&  26.764&  0.952&  0.062&  0.015&  26.341&  0.939&  0.071&  0.007& 31.600& 0.976&0.030\\ 
         Ours&  \textbf{0.011}&  \textbf{29.097}&  \textbf{0.962}&  \textbf{0.052}&  \textbf{0.010}&  \textbf{29.749}&  \textbf{0.962}&  \textbf{0.037}&  \textbf{0.006}& \textbf{33.024}& \textbf{0.980}&\textbf{0.021}\\ \hline
    \end{tabular}%
    }
    \label{table1}
\end{table}

\begin{figure}[!htbp]
  \centering
  \includegraphics[width=\textwidth]{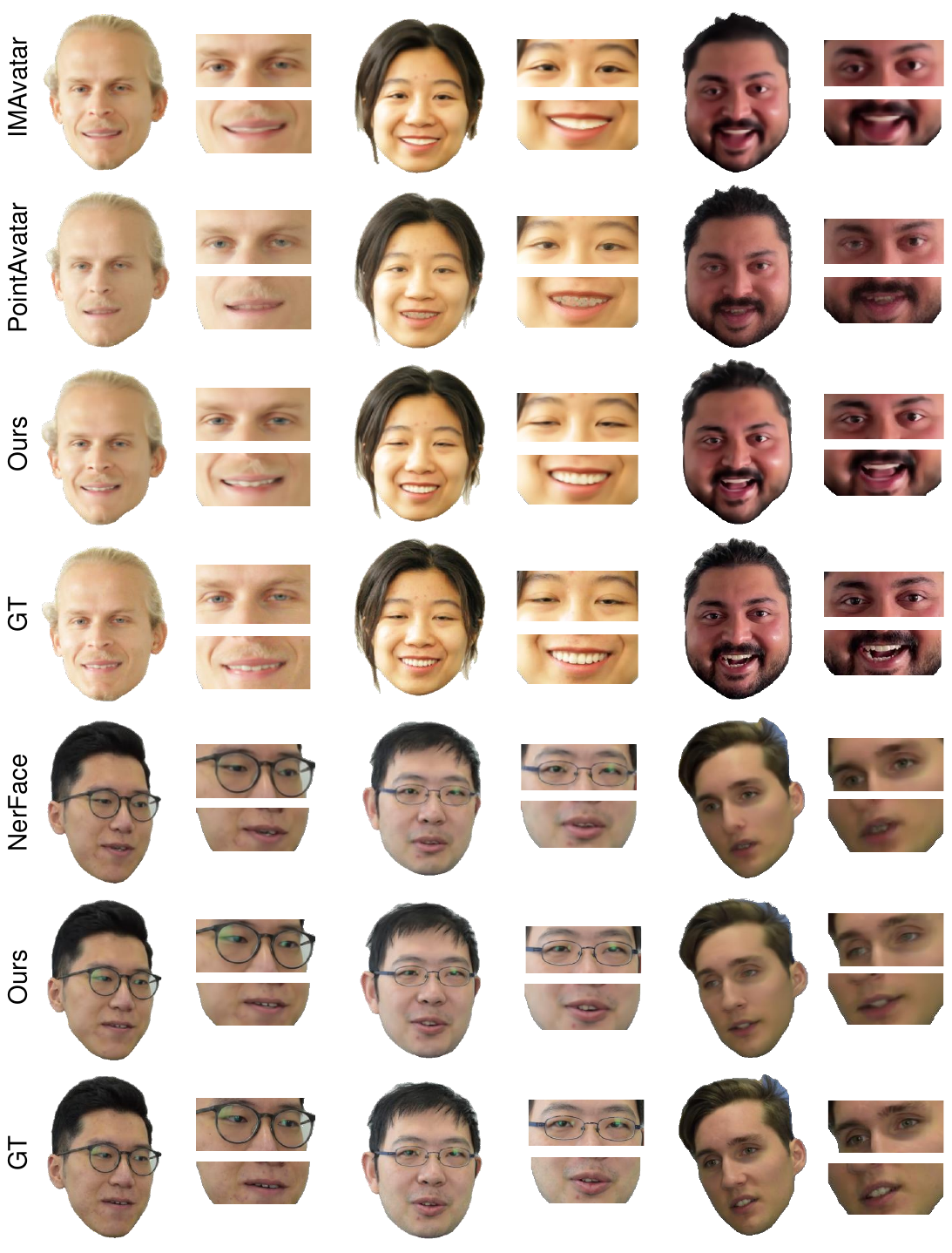}
  \caption{
    Qualitative comparisons with SOTA methods. Starting from the top left and moving down to the bottom right, each represents subjects 1, 2, 3, 4, 5, and 6, respectively. Our method demonstrates better reconstruction results and captures detailed expressions and facial structures, particularly in the eye and mouth regions.
  }
  \label{fig4}
\end{figure}

\subsection{Comparison with SOTA Methods}
\label{subsec:sota_comp}
We compared our method with three state-of-the-art (SOTA) head reconstruction methods: 1) NerFace\cite{Gafni2021DynamicNR}, which learns a dynamic neural radiance field\cite{Mildenhall2020NeRFRS} conditioned on expression and pose parameters derived from 3DMM, 2) IMAvatar\cite{Zheng2022IMA}, which is a morphing-based approach that utilizes neural implicit surfaces with learned blendshapes from diverse expressions and poses, 3) PointAvatar\cite{Zheng2023PointAvatarDP}, which constructs a facial model through the optimized deformation of point clouds. Note that for PointAvatar, the maximum number of point clouds was set to ensure similar memory usage to our model.

Qualitative comparisons on all 6 subjects are presented in Fig.~\ref{fig4}. While all methods achieve visually pleasing reconstructions, ours surpasses them by capturing finer facial details. Specifically, IMAvatar\cite{Zheng2022IMA} and PointAvatar\cite{Zheng2023PointAvatarDP} show rendering outputs with less detail. This is likely because they use a coarse 3D head template,
resulting in less effectiveness at capturing subtle nuances of facial expressions, especially on eye and mouth regions. Moreover, the results from PointAvatar show dotted artifacts due to its fixed point size. In contrast, our method employs mesh deformation to create refined geometric representations, enabling it to capture intricate changes in expression and facial structure. NerFace\cite{Gafni2021DynamicNR} uses template-based deformation for expressions, limiting its ability to represent distinct expression details across different facial regions. Consequently, it shows difficulties in accurately depicting subtle facial movements. Additionally, since it does not rely on geometric prior of 3DMM-based template, it struggles with accurately reconstructing head geometry. We also present the quantitative results in Table \ref{table1}. Note that, except for Subject 1 (whose shape is close to the template), our method outperforms others since we utilize geometric refinement.

\begin{table}[b]
    \centering
\caption{Ablation study.}
\label{table2}
    \begin{tabular}{|ccccc|} \hline 
         &  $L_1\downarrow$ &  PSNR$\uparrow$&  SSIM$\uparrow$& LPIPS$\downarrow$\\ \hline 
         w/o Per. + Smo.&  0.0265&  23.507&  0.9149& 0.0781
\\ 
         w/o Smo.&  0.0268&  23.5162&  0.9139& \textbf{0.078}\\ 
 Full model& \textbf{0.0259}& \textbf{23.7638}& \textbf{0.9151}&0.0782\\ \hline
    \end{tabular}
\end{table}

\subsection{Ablation study}
\begin{figure}[tb]
  \centering
  \begin{subfigure}{.58\textwidth}
    \centering
    \includegraphics[width=\linewidth]{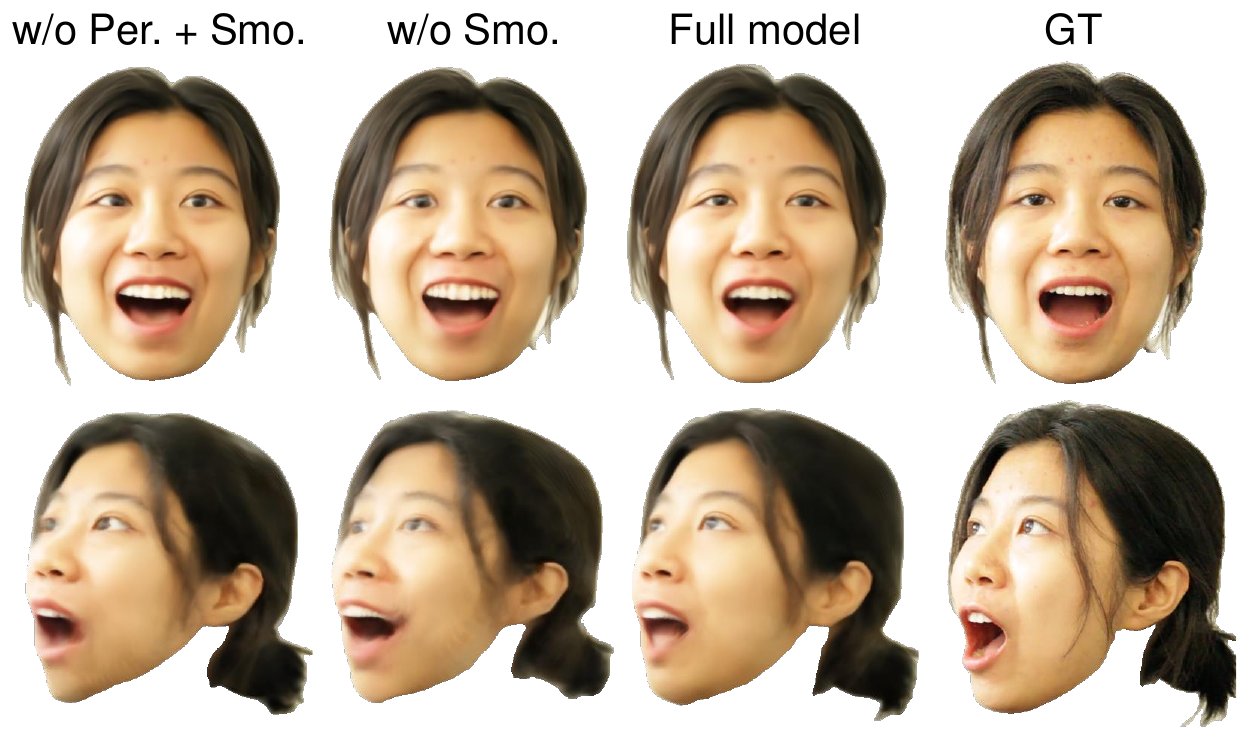}
    \caption{Ablation study}
    \label{fig5-a}
  \end{subfigure}%
  \begin{subfigure}{.42\textwidth}
    \centering
    \includegraphics[width=\linewidth]{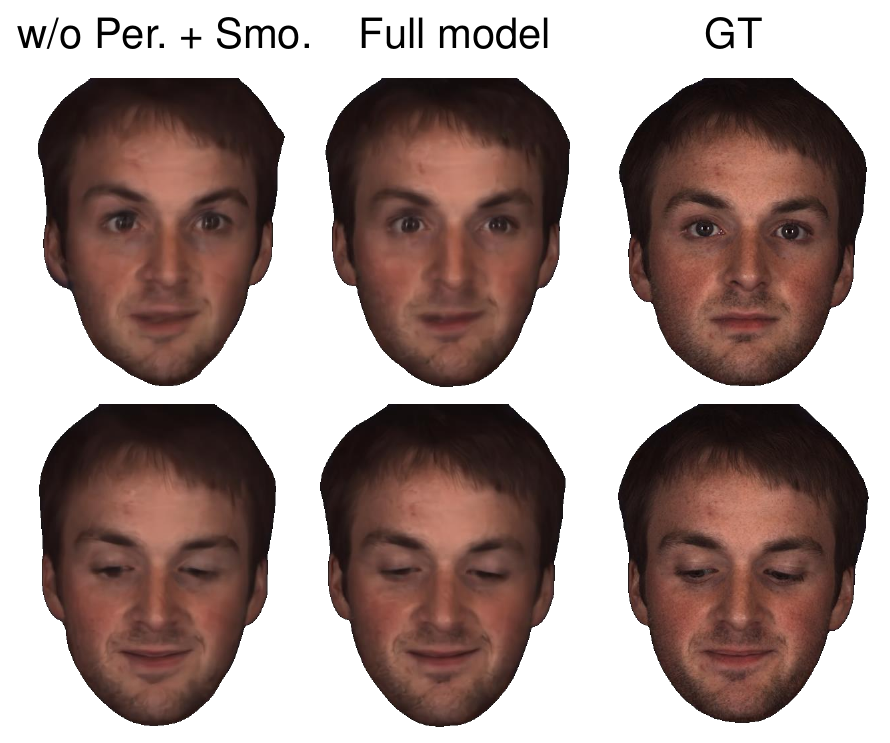}
    \caption{Test on groundtruth geometry}
    \label{fig5-b}
  \end{subfigure}
  \caption{
  \textbf{(a)} Each row represents different frames. `Per.' denotes mesh perturbation and `Smo.' denotes Laplacian smoothing. Applying both enhances the depiction of subtle facial movements. \textbf{(b)} The same abbreviations as in \textbf{(a)} are used. The model without both perturbation and smoothing demonstrates limitations in accurately capturing the facial details, and shows artifacts around the eye region, while applying the proposed method enhances the depiction of subtle facial movements.
  }
  \label{fig5}
\end{figure}

In this section, we conduct an ablation study to investigate the impact of the proposed mesh perturbation and Laplacian smoothing, both quantitatively (Table \ref{table2}) and qualitatively (Fig. \ref{fig5} for two individuals). While our base framework itself yields plausible results, as expected, the accuracy of the geometry prediction influences the quality of renderings. When the perturbation is directly applied to the mesh, high-frequency noise is introduced and the rendering output deteriorates in the detailed parts of the facial areas during extreme expressions and poses. In contrast, by employing the 2.5D mesh restricting Laplacian smoothing to the displacement to prevent overly flattening the features, our full model (including the second NeRF training) is capable of depicting subtle facial movements accurately under extreme expressions and poses. It is clearly seen that the direction of the eyes and the shape of the mouth are expressed appropriately when both perturbation and smoothing are applied.

\subsection{Comparison with Groundtruth 3D shape}
To evaluate the generated refined meshes, we assess our method on a subject featuring groundtruth geometry. The BP4D-Spontaneous dataset\cite{Zhang2013AHS,Zhang2014BP4DSpontaneousAH} is a 3D video database capturing spontaneous facial expressions. The groundtruth meshes are provided, having been obtained through passive stereo
photogrammetry. We conduct an experiment on a subject within the database, using one video for training and a different video from the same subject as the test set. The results are provided in Table \ref{table3} and Fig.~\ref{fig5-b}. Our proposed method, despite perturbing the initial mesh, which results in limited differences due to resolution constraints, more closely resembles the ground truth mesh. Additionally, as demonstrated by other metrics and the rendered results, even a slight but accurate capture of geometry may significantly influence the reconstruction outputs.

\begin{table}[t]
    \centering
\caption{Test on subject with groundtruth geometry.}
\label{table3}
    \begin{tabular}{|cccccc|} \hline 
         &  $L_1\downarrow$&  PSNR↑&  SSIM↑&  LPIPS ↓& ${L}_{2} \downarrow$ (mesh)\\ \hline 
         w/o Per. + Smo.&  0.0188&  28.4493&  0.9316&  0.1534& 0.0586
\\  
         Full model&  \textbf{0.0176}&  \textbf{28.9228}&  \textbf{0.9355}&  \textbf{0.1471}& \textbf{0.0581}\\ \hline
    \end{tabular}
\end{table}

\subsection{Visualization of Perturbation}
\begin{figure}[tb]
  \centering
  \includegraphics[width=0.8\textwidth]{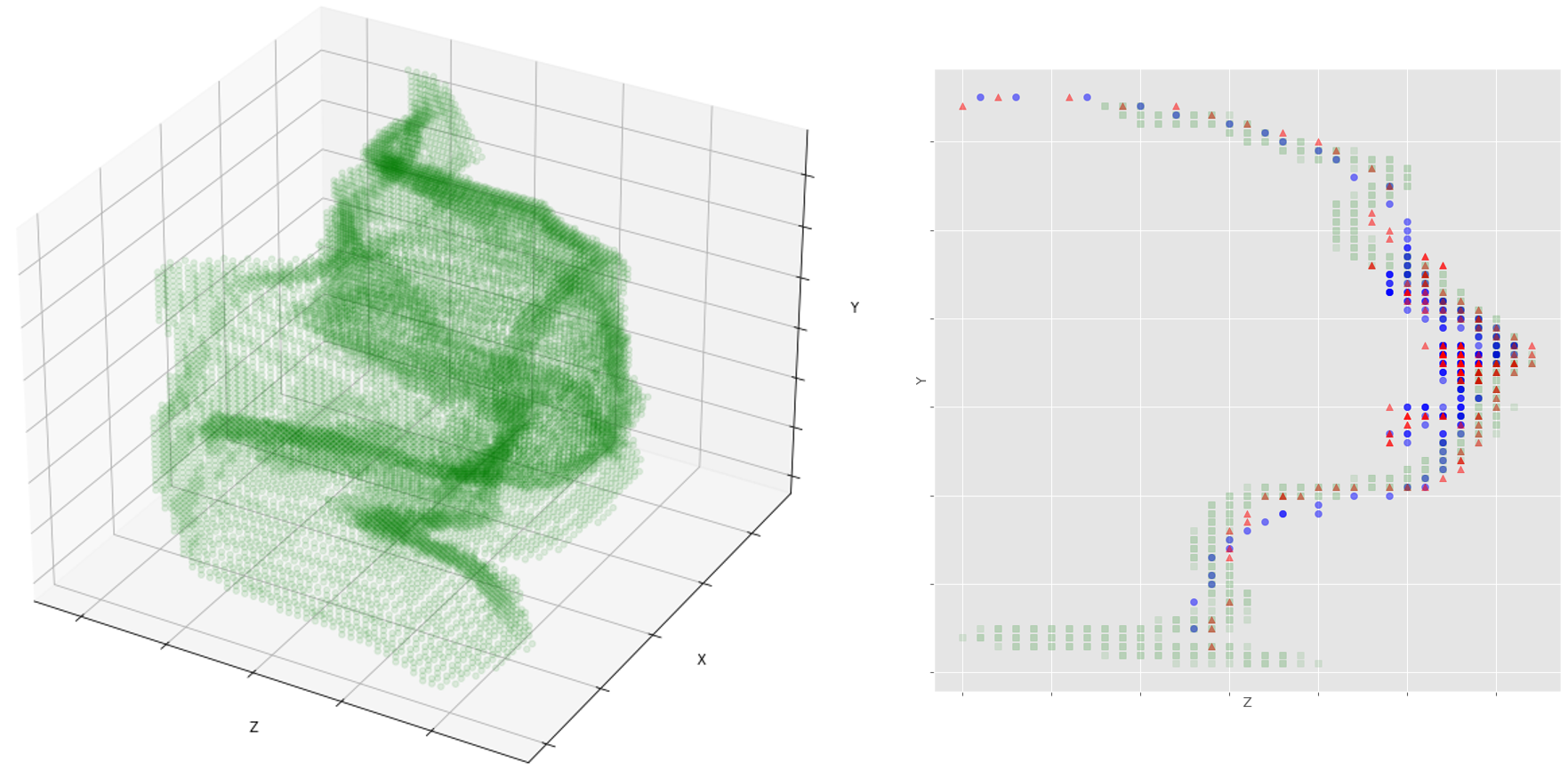}
  \caption{
    NeRF-based refinement. Left: We construct a 2.5D from the density field in NeRF. Right: Cross-section on Y-Z plane to show the perturbation, with green dots representing the 2.5D surface points, blue dots the template mesh vertices, and red dots the perturbed locations.
  }
  \label{fig6}
\end{figure}

As discussed in Sec.~\ref{subsec:perturbation}, we construct a 2.5D mesh for a smoother surface than the density field's level sets,  convert it into an SDF, and perturb the initial vertices towards the surface. In Fig. ~\ref{fig6}, the constructed 2.5D mesh and its cross-section on the Y-Z plane are visualized to demonstrate the perturbation. We can observe that the initial vertices are perturbed towards the surface.

\section{Conclusion}
\label{sec:conclusion}
We presented a geometry-guided NeRF for head avatar, based on refining a DECA 3DMM. We show in our experiments, that while NeRF constructed from an embedding of the latent code attached to vertices of a template mesh to a bounding box already exhibits superior rendering in commonly-used metrics compared with the state-of-the-art methods in the same category, further improvements are possible by perturbing the template geometry based on the density field of the NeRF itself.

\subsection{Limitations and future work}
\label{sec:limitations}
Although our tests are restricted to a specific 3DMM (DECA), and a specific way of refining the geometry (height field-based signed distance function), we expect other nonlinear 3DMMs and other geometry refinements, such as 3DGS\cite{Kerbl20233DGS} would work within our framework by replacing these two components. We also envision the bootstrapping of NeRF-based geometry for refining NeRF can be applicable to other scenes, such as full-body animation.



%
%
\bibliographystyle{splncs04}
\bibliography{main}
\end{document}